\documentclass[%
 aip,
 rsi,%
 reprint,
 amsmath,amssymb,
floatfix,
]{revtex4-1}

\usepackage{graphicx}% Include figure files
\usepackage{epstopdf}
\usepackage{dcolumn}% Align table columns on decimal point
\usepackage{bm}% bold math
\usepackage[breaklinks]{hyperref}% add hypertext capabilities
\usepackage[normalem]{ulem}

\newcommand{\ie}{\textit{i.e.}, }
\newcommand{\eg}{\textit{e.g.}, }
\newcommand{\etal}{\textit{et al}}

\newcommand{\tth}{\mathrm{th} }
\newcommand{\tEIT}{\mathrm{EIT} }

\begin{document}

%\preprint{APS/123-QED}

\title{Laser Pulse Sharpening with Electromagnetically Induced Transparency in Plasma}
\author{Kenan Qu}
\author{Nathaniel J. Fisch}%
\affiliation{%
Department of Astrophysical Sciences, Princeton University, Princeton, New Jersey 08544, USA
}%
%\affiliation{%
%Princeton Plasma Physics Laboratory, Princeton University, Princeton, %New Jersey 08543, USA
%}%

\date{\today}% It is always \today, today,
 % but any date may be explicitly specified

\begin{abstract}
We propose a laser-controlled plasma shutter technique to generate sharp laser pulses using a process analogous to electromagnetically induced transparency in atoms. The shutter is controlled by a laser with moderately strong intensity, which induces a transparency window below the cutoff frequency, and hence enables propagation of a low frequency laser pulse. Numerical simulations demonstrate that it is possible to generate a sharp pulse wavefront (sub-ps) using two broad pulses in high density plasma. 
The technique can work in a regime that is not accessible by plasma mirrors when the pulse pedestals are stronger than the ionization intensity. 
\end{abstract}

% PACS, the Physics and Astronomy Classification Scheme.
\maketitle

%\tableofcontents

\section{Introduction}
Short laser pulses are useful both because they can resolve transient phenomena and because they can reach higher intensities.
The quest for a shorter pulse duration and higher signal-to-noise ratio of a laser pulse using lower-cost experimental techniques has stimulated a rich study in both solid-state lasers and plasma compressors~\cite{PRL-Malkin1999, ExpPRL2005-Suckewer, PoP-SBS-2006, EXPPRL2010-SBS}.
%Plasma optics aiming at manipulating light have recently gained increasingly attention due to the high heat tolerance of plasma. 
In particular, plasma amplifiers based on Raman or Brillouin scattering, in which a pump pulse continuously deposits its energy into a sharp seed pulse, promise to produce exawatt laser pulses~\cite{PRL-Tsidulko2002, PRE-Yampolsky2004}. 
However, experimental realizations of such schemes are inhibited by the difficulty of preparing a sharp laser seed~\cite{PRL_plasmaseed}.  

Currently, short pulses are usually generated with a mode-locked laser, but they suffer from poor contrast ratios. Optical parametric amplifiers (OPA) produce high quality pulses, but they require a more complicated phase-matching condition. 
For improving the temporal contrast ratio of short laser pulses, a commonly used technique is plasma mirrors (PM)~\cite{Julia11PM, PM2007}. 
A PM is made of a foil or glass target which, when ionized by a strong laser field, forms a layer of overdense plasma and abruptly reflects the laser pulse. Since the low intensity laser ``prepulse'' is not sufficiently strong to induce high plasma density, it is transmitted through the PM. 
However, PM only works with an ultrashort laser with a decent initial contrast. It can only suppress the pedestals that are below the ionization intensity, which is typically $10^{14} \, \mathrm{W\,cm}^{-2}$. It also requires that the pulse be shorter than a few picoseconds; otherwise the sharp interface will be ruined by plasma expansion. Ionization costs a significant amount of the pulse energy and leaves a region that needs mechanical scanning or reconstruction. 
Therefore, preparing a sharp and clean laser pulse remains a challenge and continues to be the subject of active research.

Here, we propose an alternative scheme of generating sharp wavefronts by using two broad counter-propagating pulses with different frequencies in a high-density plasma slab. It uses a high-frequency pump laser to control the transmittance of a low-frequency seed laser in a high-density plasma slab---the plasma slab only abruptly lets the seed laser transmit when the pump intensity exceeds a threshold value. This abruptness creates a sharp wavefront in the seed transmission. For nonrelativistic seed, the timing of the shutter is controlled solely by the strong pump and it does not depend on the seed intensity. The sharpness of the wavefront depends on the frequencies of the pump laser and the plasma, regardless of initial duration or contrast ratio of the seed laser. Since the sharp wavefront is conditionally transmitted, it avoids the issue of plasma expansion or density fluctuation. 

The mechanism of the proposed optical shutter is based on an analog of electromagnetically-induced transparency (EIT)~\cite{EIT_PRL_1990, EIT_PhysT, EIT_Agarwal} in atoms. Pump laser above threshold intensity can also induce a transparency window in a high density plasma for a seed laser that is below the ``cut-off'' frequency~\cite{Harris_PRL_1996}. More comprehensively studies involving both the Stokes and anti-Stokes waves were conducted by Matsko and Rostovtsev~\cite{matskoPRE1998}, and Gordon \etal~\cite{gordonPoP20001, gordonPoP20002}. Our proposal makes use of the unique threshold behavior of EIT, which allows transforming gradual variation of optical intensity into an abrupt transmittance. Susceptible to the optical nonlinearity, transmission of the seed beam then features a steep wavefront.

\section{Theoretical model}
\begin{figure}[hpt]
	\includegraphics[width=0.35\textwidth]{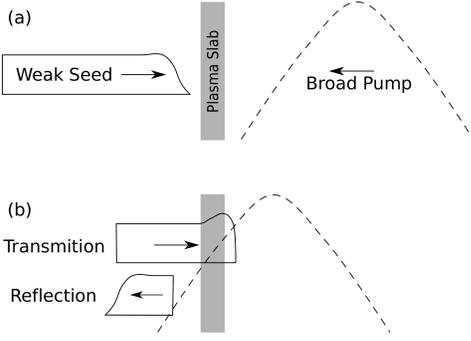}
	\caption{Schematics of the optical shutter using two broad pulses with a high-density plasma slab. (a) A nonsharpened seed below the plasma ``cut-off'' frequency gets constantly reflected in absence of pump in plasma slab. (b) When pump intensity exceeds the threshold value, the seed is abruptly let through creating a sharp wavefront.  \label{schematics}}
\end{figure}

The optical shutter, as illustrated in Fig.~\ref{schematics}, includes a high-density plasma slab with resonance frequency $\omega_{pe}$ and two counter-propagating lasers of the same polarization---a pump laser with frequency $\omega_0$ and a seed laser with frequency $\omega_1$. The laser frequencies are chosen such that 
\begin{equation}
	\omega_0-\omega_{pe} < \omega_1 < \omega_{pe} < \omega_0.
\end{equation}
When the pump intensity inside the plasma slab is low, the plasma slab reflects the seed beam which is below the ``cut-off'' frequency. As the pump intensity exceeds a threshold value, an EIT window arises between $\omega_0-\omega_{pe}$ and $\omega_{pe}$, hence the plasma slab becomes transparent for the seed beam. The seed beam is abruptly let through, yielding a sharp wavefront. The duration of the pump beam has to be sufficiently long so that the EIT window remains open before the seed wavefront exits.  During propagation, the sharpened seed pulse, upon continued interaction with the pump, can be amplified and further steepened through a Raman-backscattering-like process. 

Analogous to an atomic system, EIT in plasma arises from an interference effect enabled by the pump-driven plasma oscillation. 
When a laser beam propagates near the plasma surface, the electromagnetic wave drives electron motion in plasma. In the non-relativistic regime, electrons respond to the laser field instantaneously, \ie for laser frequency below the plasma frequency $\omega_{pe}$, the electrons oscillate at the laser frequency. In this case, the electron polarization has the same frequency but is out of phase with the incoming laser and hence they destructively interfere, leading to reflection of the laser field. 
However, a transparency window can be induced if a pump laser is applied simultaneously. The pump and probe waves beat and their ponderomotive force produces a Langmuir wave, or a density ripple. The density ripple, driven at the difference of the applied laser frequencies, perturbs the electron polarization. Importantly, if the laser frequency difference is smaller than $\omega_{pe}$, the phase of the perturbation opposes the primary electron polarization. This process induces a transparency window below the ``cut-off'' frequency and hence is called electromagnetically-induced transparency.

The propagation and amplification of the seed beam can be analyzed using the dispersion relation. We model the coupling of the laser waves and the Langmuir wave in cold plasma with the conventional coupled three-wave equations 
\begin{align}
	&(\partial_{tt} - c^2 \partial_{zz} + \omega_{pe}^2) \bm{A}_0 =  -\omega_{pe}^2 \frac{n}{\bar{n}} \bm{A}_1 , \label{1} \\
	&(\partial_{tt} - c^2 \partial_{zz} + \omega_{pe}^2) \bm{A}_1 = -\omega_{pe}^2 \frac{n}{\bar{n}} \bm{A}_0, \label{2} \\
	&(\partial_{tt} + \omega_{pe}^2) \frac{n}{\bar{n}} = c^2 \partial_{zz}(\bm{A}_0\cdot\bm{A}_1),  \label{3} 
\end{align}
where $\bm{A}_0$ and $\bm{A}_1$ are the vector potentials of the pump laser and seed laser normalized to $e/m_ec$, respectively, with $e$ being the natural charge, $m_e$ being the electron mass, and $c$ being the speed of light. The electron density variation $n/\bar{n}$ describes the Langmuir wave which oscillates at frequency $\omega_{pe}= \sqrt{\bar{n}e^2/(\epsilon_0m_e)}$, with $\epsilon_0$ being the permittivity of vacuum. We first assume a weak initial seed and a nondepleted pump by neglecting dynamics of the pump [Eq.~(\ref{1})]. The dispersion relation can then be exactly derived in the frequency regime~\cite{matskoPRE1998, gordonPoP20001} 
\begin{equation}\label{disp}
  \omega_1^2 - c^2 \bm{k}_1^2  = \omega_{pe}^2 - f c^2 (\bm{k}_0-\bm{k}_1)^2, 
\end{equation}
where $f= A_0^2/[1 - (\Delta\omega/\omega_{pe})^2]$ with $\Delta\omega=\omega_0-\omega_1$ being the two-photon detuning. 
Note that one cannot make $\Delta\omega =\omega_{pe}$ because it invalidates negligence of electron thermal velocity and damping. Compared with the normal dispersion relation of electromagnetic waves in plasma, Eq.~(\ref{disp}) includes an extra term which depends on the pump intensity.

\begin{figure}[htp]
	\includegraphics[width=0.35\textwidth]{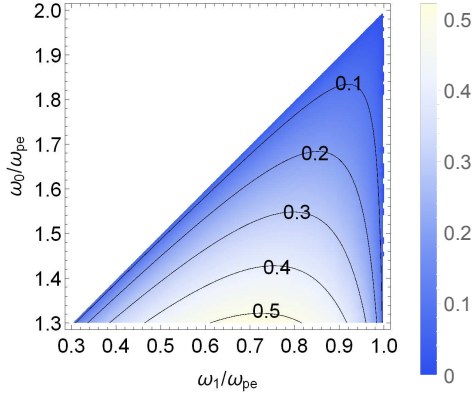}
	\caption{(Color Online.) Contour plot of the threshold value of pump amplitude $A_\tth$ for EIT as a function of pump frequency $\omega_0$ and seed frequency $\omega_1$ normalized to plasma frequency $\omega_{pe}$. In presence of a pump, EIT window (colored region) emerges at seed frequencies close to $\omega_0-\omega_{pe}$ and close to $\omega_{pe}$. Note that EIT window does not exist in the blank region on top left.   \label{pumpA}}
\end{figure}

\begin{figure}[htp]
	\includegraphics[width=0.48\textwidth]{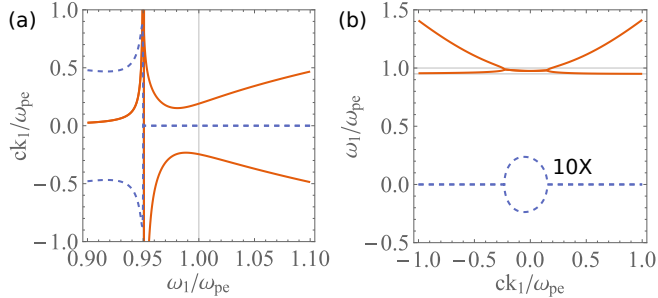}
	\caption{(Color Online.) The real (red solid) and imaginary (blue dashed) parts of the roots of the dispersion relation [Eq.~(\ref{disp})] at $A_0=0.04$. In (a), seed frequency $\omega_1$ is set as real with wavevector $k_1$ being complex; in (b), seed wavevector $k_1$ is set as real with frequency $\omega_1$ being complex. The dashed curve in (b) is multiplied by a factor of $10$ for illustration purpose. The pump frequency is $\omega_0=1.95\omega_{pe}$ and its corresponding wavevector is $ck_0=1.67\omega_{pe}$. The thin gridlines show $\omega_0$ and $\omega_0-\omega_{pe}$ for reference. An EIT window and instability emerge between $\omega_0-\omega_{pe}$ and $\omega_{pe}$. \label{figdisp}}
\end{figure}

Conditions of transparency can be found by solving the dispersion relation [Eq.~(\ref{disp})] for $\bm{k}_1$ by setting $\omega_1$ real. 
%If we limit our discussion to forward and backward scattering, \ie $\bm{k}_1 \| \bm{k}_0$, 
The solution can be expressed as 
\begin{equation}\label{eqk}
\bigg|\bm{k}_1 - \frac{f}{1-f}\bm{k}_0 \bigg| = \frac{1}{(1-f)c} \sqrt{f(\omega_0^2-\omega_1^2) - (\omega_{pe}^2-\omega_1^2)}. 
\end{equation}
The real roots of wavevector $\bm{k}_1$ lies on a circle centered at $f\bm{k}_0/(1-f)$. They become purely real so that the beam begins to propagate when the pump intensity is above the threshold value $A_\tth$, \ie
\begin{equation}
	A_0^2 \geq A_\tth^2 \equiv \bigg[1-\Big(\frac{\Delta\omega}{\omega_{pe}}\Big)^2 \bigg] \frac{\omega_{pe}^2 - \omega_1^2}{\omega_0^2-\omega_1^2} \label{13} 
\end{equation}
on the condition that $\omega_{pe}> \Delta\omega >0$; otherwise the inequality (\ref{13}) is to be reversed. We show in Fig.~\ref{pumpA} a contour plot of the threshold value of pump amplitude $A_\tth$ for EIT as a function of pump frequency $\omega_0$ and seed frequency $\omega_1$. It is evident that a lower pump frequency enables a larger bandwidth of transparency window, but it also requires a higher threshold of pump amplitude. As an example of cases with a strong pump, we plot in Fig.~\ref{figdisp}(a)  the real roots of $k_1$ assuming $\bm{k}_1 \| \bm{k}_0$. It shows that the transparency window, once opened, extends to $\omega_0-\omega_{pe}$ which is below the ``cut-off'' frequency. Its spectra width is $\omega_\mathrm{EIT}=2\omega_{pe}-\omega_0$. 

We note that the threshold value for EIT [Eq.~(\ref{13})] does not depend on the directions of wavevectors, which allows to use an arbitrary angle between the pump and seed beams. Although we focus on two counter-propagating lasers for simplicity of analysis and simulation, the flexibility might be an advantage in experiments as it avoids the issue of aligning two optical pulses.

Instabilities of propagating seed beam can be found by analyzing the root $\omega_1$ as a function of real wavevector $k_1$. Under the condition $\omega_0 \lesssim 2\omega_{pe}$, the solution to Eq.~(\ref{disp}) can be well approximated as 
\begin{align}
	\omega_1 &\cong \frac12(\omega_0 + \omega_h - \omega_{pe}) \nonumber \\
	&  \pm \frac12 \sqrt{(\omega_{pe}-\omega_0 + \omega_h)^2 - A_0^2 \frac{2\omega_{pe}^2 (\omega_0^2-\omega_{pe}^2) }{\omega_0\omega_h}}, \label{8}
\end{align}
where $\omega_h = \sqrt{\omega_{pe}^2 + c^2k_1^2}$. When the pump amplitude is above the threshold value, roots of $\omega_1$ become complex for real wavevectors $k_1$, which indicates an instability in this region. An example is shown in Fig.~\ref{figdisp}(b). From the plot, we can see the dispersion curves have two branches in the regions of large wavevector $k_1$. They, respectively, correspond to the Stokes sideband of the pump at $\omega_0-\omega_{pe}$ and a frequency up-shifted mode at $\omega_h$. The latter mode has been studied by Wilks \etal~\cite{PRL-Wilks-1988} in the scheme of flash ionization in which electromagnetic field couples to plasma wave when an overdense plasma is abruptly created. In the regions of small wavevector $k_1$, these two branch couples and lead to an instability. The maximum growth rate of instability  exists at a point with a negative wavevector and with a frequency between $\omega_h=\omega_{pe}$ and $\omega_h=\omega_0-\omega_{pe}$. Hence, backward scattering in a long plasma eventually dominates the outputs. 

The sharpened wavefront of the seed beam arises from the abrupt change of the plasma dispersion relation, since it is controlled by pump pulse intensity. When the EIT window opens, the Langmuir wave and seed beam excite each other and propagate. 
Since the dispersion relation of the seed depends on the pump intensity, we analyze the seed beam dynamics by first concentrating on the linear stage where pump intensity remains a constant. In this regime, the monochromatic seed beam has a definite wavevector $k_1$. Hence we can take the envelope approximation and write $\bm{A}_1 = \tilde{\bm{A}}_1 e^{-i(\omega_1t-k_1z)}$ and $n=\tilde{n} e^{-i[\Delta\omega t-(k_0-k_1)z]}$. Then Eqs.~(\ref{2}) and (\ref{3}) can be rewritten as
\begin{align}
	& 2i(\omega_1 \partial_t - c^2k_1 \partial_z) \tilde{\bm{A}}_1 = \omega_{pe}^2 \frac{\tilde{n}}{\bar{n}} \tilde{\bm{A}}_0 + D_1 \tilde{\bm{A}}_1, \label{9} \\
	& 2i\Delta\omega \partial_t \frac{\tilde{n}}{\bar{n}} = -c^2(k_0-k_1)^2 \tilde{\bm{A}}_0 \cdot \tilde{\bm{A}}_1 + D_n \frac{\tilde{n}}{\bar{n}}, \label{10}
\end{align}
where $D_1=\omega_{pe}^2 + c^2k_1^2 - \omega_1^2$ and $D_n = \omega_{pe}^2 - \Delta\omega^2$ describe dispersion of the waves. The dispersion terms are negligible~\footnote{The dispersion terms, when being considered to the first order perturbation, only induce oscillations of wave amplitudes at frequencies of $D_1/\omega_1$ and $D_n/\Delta\omega$, respectively. } as long as values of $\tilde{n}/\bar{n}$ and $\tilde{\bm{A}}_1$ are both smaller than $\tilde{\bm{A}}_0$. As such, Eqs.~(\ref{9}) and (\ref{10}) become the well-known coupled wave equations for describing linear stage of Stimulated Raman Scattering (SRS)~\cite{SRS_1994}. Note that an EIT process is different from SRS, although their dynamic equations have a similar form. This can be seen from the frequency relations; EIT requires a nonresonant pump  $\omega_0 < \omega_1 + \omega_{pe}$ and SRS uses a resonant pump for the maximum growth rate. Actually, SRS does not exist in a plasma of above one quarter of the critical density, where EIT is operated.
The solution to Eqs.~(\ref{9}) and (\ref{10}) has been studied extensively and can be expressed by convolution of the seed with a first order associated Bessel function. In our case, the amplified seed features an exponentially growing wavefront with growth rate $\Gamma = \mathrm{Im}(\omega_1)$. In the nonlinear stage, the growth rate begins to decrease when pump intensity begins to deplete at the seed peak.  The seed tail even stops propagating once the pump intensity falls below the threshold value. However, the seed wavefront continues to grow and gets sharpened. Its sharpness, defined by the ``rising-time'' $t_r$,  approaches the asymptotic value which is limited by the finite frequency bandwidth of the EIT window
\begin{equation}\label{tr}
t_r = \frac{1}{\omega_\tEIT} = \frac{1}{2\omega_{pe}-\omega_0}.
\end{equation}

Therefore, in order to increase the maximum seed sharpness, the plasma frequency $\omega_{pe}$ is preferably set close to half the pump frequency $\omega_0$.
In the most favorable regime with seed frequency $\omega_1\sim\omega_0/2$, the rising edge of the obtained pulse only contains a small number of optical cycles, \ie $\omega_1 t_r = 1/[2(\omega_{pe}/\omega_1 -1)]$. Note that the input seed beam can be of any sharpness and even a continuous wave. Given a long plasma, the seed intensity can continue to grow. However, the seed also suffers strong group velocity dispersion (GVD) associated with the nonlinear dispersion relation which reduces pulse sharpness. Thus, a thin plasma slab is desirable for optimal sharpness of the pulse output. Its minimum thickness is confined by seed tunneling with a characteristic length $c\Big/\sqrt{\omega_{pe}^2-\omega_1^2}$.  For higher pulse fluence, the sharpened pulse can be sent into a lower-density plasma medium for second stage amplification.

\section{Numerical simulations}

In order to demonstrate numerically this effect, we conduct full one-dimensional PIC simulations using the code EPOCH~\cite{EPOCH2015}. Two counter-propagating laser pulses are sent into a thin plasma slab with electron density $n_e=1.14\times 10^{20}  /\mathrm{cm}^3$ and correspondingly $\omega_{pe} = 2\pi\times 95.867\mathrm{THz}$. The left-going pump laser pulse has a frequency $\omega_0= 1.95\omega_{pe}= 2\pi\times 187.50\mathrm{THz}$. The right-going seed pulse has a frequency $\omega_1=0.975\omega_{pe}$ which is below plasma frequency. Both of the laser pulses have the same Gaussian shape with a half width at half maximum (HWHM) of $0.59\mathrm{ps}$ ($0.176\mathrm{mm}$ spatially).  With this set of parameters, rising time of the sharpened seed would be limited to $t_r\approx 0.1 \mathrm{ps}$. The threshold pump intensity for transparency at seed frequency is $I_\mathrm{th} = 1.7\mathrm{PW/cm}^2$ at which relativistic effect is negligible. We keep the temperature low ($10\mathrm{eV}$) to suppress Landau damping. A cell size of $4\mathrm{nm}$ is used to match the Debye length, and $300$ electrons per cell are employed to reduce the charge density fluctuation. Ion motions are ignored in the simulation. 

\begin{figure}[htp]
  \includegraphics[width=0.48\textwidth]{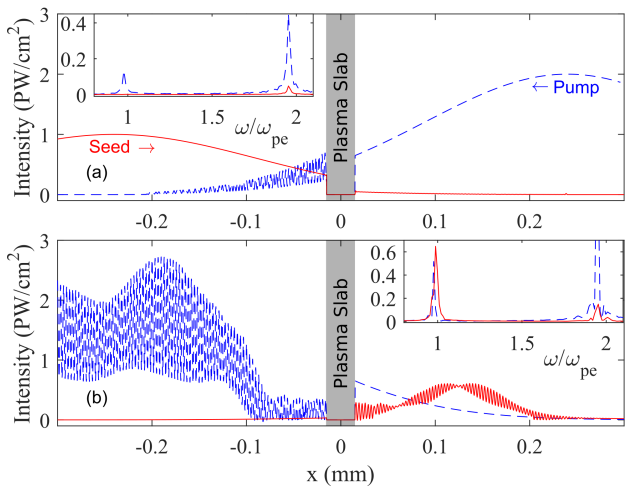}
  \caption{(Color Online.) PIC simulation of the laser pulses after (a) $t=1.8\mathrm{ps}$ and (b) $t=3.2\mathrm{ps}$. The blue dashed and red solid curves show the envelopes of the pump and seed intensity, respectively. For illustration, red curve in (a) is multiplied by a factor of $10$. The inset shows  spectra of the output (see main text for details). The weak broad seed pulse ($\mathrm{HWHM}=0.59\mathrm{ps}$) is compressed and amplified to a strong sharp pulse [$\mathrm{HWHM}=0.15\mathrm{ps}$ shown in (b)]. \label{PICSq}}
\end{figure}

We present two snapshots of the simulation results in Fig.~\ref{PICSq}. Figure~\ref{PICSq}(a) shows the pulse envelopes (blue dashed for pump which propagates towards left; and red solid for seed  which propagates towards right) at $t=1.8\mathrm{ps}$ before which the pump pulse intensity gradually grows but remains below $I_\mathrm{th}$. We indeed do not observe any seed transmission in Fig.~\ref{PICSq}(a) \footnote{Wave intensities inside the region of plasma is not shown because the phase velocity, which is required for distinguish left- and right-propagating waves, is not well defined for hybrid modes.}. The inset shows frequency components of (i) right-propagating wave in the region $x>0.03 \,\mathrm{mm}$ using red solid curve; and (ii) left-propagating wave in the region $x<-0.03 \,\mathrm{mm}$ using blue dashed curve. They are calculated using $\omega=2\pi c/\lambda$ where $\lambda$'s are obtained by Fourier transforming the transmitted signal. The inset shows three peaks---peak of the red curve at pump frequency ($1.95\omega_{pe}$) is the reflection of pump beam on the plasma surface; and peaks of the blue curve correspond to the transmitted pump and reflected seed, respectively.  They beat producing the oscillation in the region between $x=-0.2\mathrm{mm}$ and $x=-0.03\mathrm{mm}$. 
It shows no transmission at seed frequency ($0.975\omega_{pe}$).

As the interaction continues, the pump intensity grows above $I_\mathrm{th}$. The EIT window then opens and the seed enters the plasma slab. The PIC simulation result at $t=3.2\mathrm{ps}$ is shown in Fig.~\ref{PICSq}(b). We first observe a strong and short amplified seed pulse between $x=0.03\mathrm{mm}$ and $x= 0.2\mathrm{mm}$. Its peak intensity reaches about $0.7\mathrm{PW/cm}^2$ which is seven times stronger than the input seed. Its wavefront duration is strongly compressed to a HWHM of about $0.15\mathrm{ps}$ ($0.04\mathrm{mm}$ spatially). The inset show that the central frequency of the amplified seed is indeed below $\omega_{pe}$. Oscillation in the transmitted seed arises from beating with the reflected pump. We also observe strong beating signal in the transmitted pump, which confirms our theory that instability exists in both directions. 

\begin{figure}[htp]
	\includegraphics[width=0.48\textwidth]{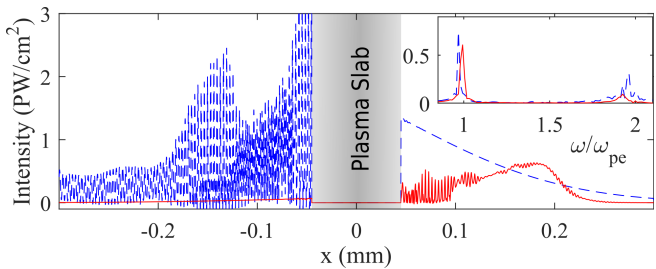}
	\caption{(Color Online.) PIC simulation results of the laser pulses $t=2.7\mathrm{ps}$ when the plasma density has a Gaussian distribution with $\mathrm{HWHM}=35.2\mu\mathrm{m}$. Other parameters are identical to Fig.~\ref{PICSq}. It shows the pump reflection is strongly suppressed by the plasma density gradient.  \label{PICGa}}
\end{figure}

In the above ``proof-of-principle'' example, we notice the relatively high reflectance of the pump beam at the surface of plasma slab. The reflected pump beam becomes a precursor and reduce the seed pulse contrast. According to Fresnel reflection equation, the reflectance is 
\begin{equation}
\mathcal{R} = \left| \frac{1- \sqrt{1-(\omega_{pe}/\omega_0)^2} }{ 1+\sqrt{1 -(\omega_{pe}/\omega_0)^2}} \right|^2 . 
\end{equation} 
For purpose of suppressing pump reflectance, higher values of $\omega_0/\omega_{pe}$ are preferred. Note, however, that inequality $\omega_0 > \omega_1+\omega_{pe}$ has to be obeyed, otherwise seed frequency falls outside the EIT window. An alternative means to suppress pump reflection is to introduce a density gradient close to the plasma surface. Functioning similar to an anti-reflection coating, the gradually varying plasma frequency can provide optical impedance matching~\cite{rayleigh1879reflection} and significantly minimize pump reflection. The corresponding PIC simulation result is shown in Fig.~\ref{PICGa} where the plasma density profile is $n_e=1.14\times 10^{20} \times e^{-(x/30\mu\mathrm{m})^2} \,\mathrm{cm}^{-3}$. The amplified seed reveals an extremely sharp wavefront with a rise time of $t_r\sim 0.1 \mathrm{ps}$, which agrees with Eq.~(\ref{tr}) very well. Importantly, its front edge does not show any broad precursors. From the inset, we find the reflected pump intensity is below $10\%$ of the amplified seed; hence, the beating is negligible. 

In order to distinguish the mechanism of pulse compression from the amplification process, we compare our results with backward Raman amplification. We run a PIC simulation with the plasma density reduced by a factor of $2$, so that $\omega_1 > \omega_{pe}$. The results show transmission of the seed does not depend on the pump intensity, as expected. The transmitted seed contains a broad wavefront that is similar to the original seed. This is in contrast to the result using EIT which suppresses the precursors. Therefore, the simulations show that the threshold behavior of EIT is essential in pulse sharpening.

\section{Conclusion}
%The proposed optical shutter can be directly applied to current backward Raman amplifiers in plasma. 

We considered propagation of a seed laser pulse in an overdense plasma controlled by a separate pump pulse. When the pump intensity reaches a threshold value, it abruptly induces an EIT window for the seed pulse, yielding a sharp wavefront of the transmitted seed pulse. During propagation, the seed pulse gets amplified by a Raman-like parametric instability. The output intensity of the sharpened seed can be comparable to pump. The sharpness of the obtained pulse is characterized by the ``rise-time'' of the pulse wavefront, \ie $t_r=1/(2\omega_{pe}-\omega_0)$.  For optimal operation, the plasma slab should be thin enough to mitigate the GVD effect, but it should also be thicker than the characteristic length of seed tunneling ($c\Big/\sqrt{\omega_{pe}^2-\omega_1^2}$).

The threshold of pump amplitude, as expressed in Eq.~(\ref{13}), increases at a larger plasma frequency $\omega_{pe}$ which depends on plasma density and temperature. In a laser-induced plasma channel, finite plasma temperature often deteriorates plasma wave by increasing plasma frequency and Landau damping. These deleterious effects result in an increased threshold value and decreased growth rate of instability. Therefore, for given pump and seed pulses, random density fluctuation and finite temperature reduce the peak intensity of the obtained pulse. Nevertheless, the wavefront sharpness of the obtained pulse is immune to these deleterious effects. Since the seed frequency is below the plasma frequency, forward or side Raman scattering of the seed is automatically suppressed. Of interest in high-density plasmas is the Brillouin scattering of the pump beam due to the ion acoustic wave whose growth rate is approximately $\omega_{pi}=\omega_{pe}\sqrt{m_e/m_i}$ where $m_e/m_i$ is the mass ratio of electron to ion. Using heavier ions would reduce the growth rate of this interaction, so that the ion wave would not reach significant amplitude within a thin plasma slab.

Note that one should not confuse EIT with relativistic transparency (RT)~\cite{RIT_2012}in plasmas. In RT, electrons, driven by superintense lasers, reach near light-speed and thus increase in mass. It slows the electron motion so that plasma can no longer shield the electromagnetic wave, and hence transparency is induced. Although RT can also be used for pulse sharpening, it requires that the pulse itself be very intense (above $10^{18}-10^{19} \,\mathrm{W \,cm}^{-2}$).  Strong ponderomotive forces cause plasma expansion and the associated Doppler effect limits the pulse sharpness. 
In contrast, EIT arises from interference within the plasma wave interacting with different laser fields. This allows the use of weak pump lasers whose intensity can be well below relativistic regime, \eg $ 10^{15} \,\mathrm{W\,cm}^{-2}$.

\section{Acknowledgments}
This work was supported by NNSA Grant No. DE-NA0002948, and AFOSR Grant No. FA9550-15-1-0391. The EPOCH code was developed as part of the UK EPSRC 300 360 funded project EP/G054940/1.

\Urlmuskip=0mu plus 1mu\relax
\bibliography{PlasmaEIT-ref}

\end{document}